\documentclass[fleqn, usenatbib, useAMS]{mnras}

\usepackage{newtxtext,newtxmath}
\usepackage[T1]{fontenc}
\usepackage{ae,aecompl}
\usepackage[utf8]{inputenc}

\usepackage{graphicx}	% Including figure files
\usepackage{amsmath}	% Advanced maths commands
\usepackage{bm}		    % Bold maths symbols, including upright Greek

\newcommand{\Tr}{\mathrm{Tr}}

\newcommand{\skm}{\,s\,km$^{-1}$\;}
\newcommand{\kms}{\,km\,s$^{-1}$\;}
\newcommand{\hmpc}{\,$h$\,Mpc$^{-1}$\;}

\newcommand{\poned}{$P_{\mathrm{1D}}$\;}

\defcitealias{faucher-giguere_direct_2008}{FG08}
\title[$P_{\mathrm{1D}}$ -- KODIAQ, SQUAD \& XQ-100]{Optimal 1D Ly$\alpha$ Forest Power Spectrum Estimation -- II. KODIAQ, SQUAD \& XQ-100}

\author[N. G. Kara{\c c}ayl{\i} et al.]
{Naim G{\" o}ksel Kara{\c c}ayl{\i}$^{1}$, 
Nikhil Padmanabhan$^{1, 2}$,
Andreu Font-Ribera$^{3, 4}$,
Vid Ir{\v s}i{\v c}$^{5}$, 
Michael Walther$^{6}$,
\newauthor
David Brooks$^{4}$, 
Enrique Gazta{\~{n}}aga$^{7, 8}$,
Robert Kehoe$^{9}$,
Michael Levi$^{10}$, 
Pierros Ntelis$^{11}$, 
\newauthor
Nathalie Palanque-Delabrouille$^{6}$,
Gregory Tarl{\' e}$^{12}$
\\
$^{1}$Department of Physics, Yale University, New Haven, CT, USA\\
$^{2}$Department of Astronomy, Yale University, New Haven, CT, USA\\
$^{3}$Institut de F\'isica d'Altes Energies (IFAE), The Barcelona Institute of Science and Technology, 08193 Bellaterra (Barcelona), Spain\\
$^{4}$Department of Physics and Astronomy, University College London, London, UK\\
$^{5}$Kavli Institute for Cosmology, Department of Physics, University of Cambridge, Madingley Road, Cambridge CB3 0HA, UK\\
$^{6}$IRFU, CEA, Université Paris-Saclay, F-91191 Gif-sur-Yvette, France\\
$^{7}$Institut d'Estudis Espacials de Catalunya (IEEC), E-08034 Barcelona, Spain\\
$^{8}$Institute of Space Sciences (ICE, CSIC), Campus UAB, Carrer de Can Magrans, s/n, E-08193 Barcelona, Spain\\
$^{9}$Department of Physics, Southern Methodist University, 3215 Daniel Ave., Dallas, TX 75205, USA\\
$^{10}$Lawrence Berkeley National Laboratory, 1 Cyclotron Road, Berkeley, CA 94720, USA\\
$^{11}$Aix-Marseille Univ, CNRS/IN2P3, CPPM, Marseille, France\\
$^{12}$Department of Physics, University of Michigan, Ann Arbor, MI 48109, USA
}

\date{Accepted XXX. Received YYY; in original form ZZZ}

\pubyear{2021}

\begin{document}
\label{firstpage}
\pagerange{\pageref{firstpage}--\pageref{lastpage}}
\maketitle

\begin{abstract}
We measure the 1D Ly\,$\alpha$ power spectrum \poned from Keck Observatory Database of Ionized Absorption toward Quasars (KODIAQ), The Spectral Quasar Absorption Database (SQUAD) and XQ-100 quasars using the optimal quadratic estimator. 
We combine KODIAQ and SQUAD at the spectrum level, but perform a separate XQ-100 estimation to control its large resolution corrections in check.
Our final analysis measures \poned at scales $k<0.1$\skm between redshifts $z=$ 2.0 -- 4.6 using 538 quasars.
This sample provides the largest number of high-resolution, high-S/N observations; and combined with the power of optimal estimator it provides exceptional precision at small scales.
These small-scale modes ($k\gtrsim 0.02$\skm), unavailable in Sloan Digital Sky Survey (SDSS) and Dark Energy Spectroscopic Instrument (DESI) analyses, are sensitive to the thermal state and reionization history of the intergalactic medium, as well as the nature of dark matter. 
As an example, a simple Fisher forecast analysis estimates that our results can improve small-scale cut off sensitivity by more than a factor of 2.

% We find good agreement with previous measurements at the largest scales $k\lesssim0.003$\,s\,km$^{-1}$, even though this set is still limited by statistics, continuum errors and DLAs for these scales
\end{abstract}


%%%%%%%%%%%%%%%%%%%%%%%%%%%%%%%%%%%%%%%%%%%%%%%%%%

%%%%%%%%%%%%%%%%% BODY OF PAPER %%%%%%%%%%%%%%%%%%

\section{Introduction}
The Ly\,$\alpha$ forest technique can map out the matter distribution in vast volumes and at small scales ($r\lesssim 1$\,Mpc) between $2 \lesssim z \lesssim 5$. 
At these redshifts, the structure formation is mildly non-linear, but the physics of the Ly\,$\alpha$ forest is further enriched by the thermal state and reionization history of the intergalactic medium (IGM) \citep{hui_equation_1997, gnedin_probing_1998}. 
The line-of-sight flux power spectrum \poned has been at the frontier of constraining new physics including IGM thermal evolution \citep{boera_revealing_2019, walther_new_2019}, neutrino masses \citep{croft_cosmological_1999, seljak_can_2006, palanque-delabrouille_constraint_2015,palanque-delabrouille_neutrino_2015, yeche_constraints_2017} and the nature of dark matter \citep{boyarsky_lyman-$upalpha$_2009, viel_warm_2013, baur_lyman-alpha_2016, irsic_first_2017, garzilli_lyman-_2019}. 

Two categories of \poned data sets have emerged over the years. The first category contains thousands of low- to medium-resolution spectra obtained by the Extended Baryon Oscillation Spectroscopic Survey (eBOSS) \citep{dawson_sdss-iv_2016} and its predecessors; and the corresponding \poned estimates \citep{mcdonald_ly$upalpha$_2006, palanque-delabrouille_one-dimensional_2013, chabanier_one-dimensional_2019}. 
The upcoming Dark Energy Spectroscopic Instrument (DESI) \citep{levi_desi_2013, desi_collaboration_desi_2016} aims to obtain approximately one million Ly\,$\alpha$ quasar spectra. Such large sample sizes can probe large scales to constrain cosmology and neutrino masses, but these data sets are limited by noise and resolution at small scales.
The second category contains tens to hundreds of high-resolution, high-S/N spectra obtained by various spectrographs including the High-Resolution Echelle Spectrograph  \citep[HIRES]{vogt_hires_1994}, the Ultraviolet and Visual Echelle Spectrograph  \cite[UVES]{dekker_design_2000} and X-Shooter spectrograph \citep{vernet_x-shooter_2011}. \poned estimates in this category often have been limited to their respective data sets with 10--100 spectra \citep{croft_power_1999, mcdonald_observed_2000, kimPowerSpectrumFlux2004, viel_warm_2013, walther_new_2017, irsic_lyman_2017,yeche_constraints_2017, boera_revealing_2019, day_power_2019}. The large-scale modes ($\sim 10$\,Mpc) are poorly measured due to large sample variance from the small numbers of quasars in this category, but these spectra can probe extremely small scales ($\sim 100$\,kpc) which are crucial to constrain thermal state of the IGM and non-standard dark matter models.

In this work, we measure the small-scale \poned from the largest sample of high-S/N quasars using a combination of three public releases \citep{lopez_xq-100_2016, omeara_second_2017, murphy_uves_2019}. This combined sample is seven times larger than \citet{walther_new_2017}, but requires attention when combining the different data sets.
%%In fact, we find that XQ-100 is inconsistent with the other two and remove it from our final conservative estimate which uses 464 unique quasars.
%%NP: this should be in the summary
We use the optimal quadratic estimator formalism and pipeline described in \cite{karacayli_optimal_2020} to measure \poned. This approach has a number of advantages: by working in pixel space,
it is unbiased by gaps in the spectra, allows weighting by both the pixel-level pipeline noise as well as accounting for sample variance from intrinsic Ly\,$\alpha$ correlations, and naturally allows combining very different data sets (with different pixel spacings, resolutions etc).
As we demonstrate below, the large data set and pipeline result in significant improvements to the precision with which we measure \poned.

This paper is organized in seven sections. Section~\ref{sec:data} describes the spectra and the preprocessing steps we take. Section~\ref{sec:method} details our method with a summary of the pipeline and mean flux measurement. We also review the quadratic estimator and validate the pipeline on simulated data here. Our results and a discussion on systematics of damped Ly\,$\alpha$ absorbers, continuum errors and metal contamination can be found in Section~\ref{sec:results}. We reflect on our results and their statistical power in Section~\ref{sec:discussion}. Finally, we summarize in Section~\ref{sec:summary}. Data availability is in Section~\ref{sec:data_avail}.

\section{Data}
\label{sec:data}
We use three publicly available data sets in this work:
\begin{itemize}
    \item Keck Observatory Database of Ionized Absorption toward Quasars (KODIAQ) Data Release 2 (DR2) \citep{lehner_galactic_2014,omeara_first_2015,omeara_second_2017} is observed by HIRES  \citep{vogt_hires_1994} on the Keck I telescope.
    \item The Spectral Quasar Absorption Database (SQUAD) DR1 \citep{murphy_uves_2019} is observed by UVES  \citep{dekker_design_2000} on the European Southern Observatory’s Very Large Telescope (VLT).
    \item XQ-100 is observed using the X-Shooter spectrograph \citep{vernet_x-shooter_2011} under the European Southern Observatory Large Programme “Quasars and their absorption lines: a legacy survey of the high-redshift universe with VLT/XSHOOTER” \citep{lopez_xq-100_2016}.
\end{itemize}

The availability of large, high-resolution, high-S/N spectra of KODIAQ and SQUAD also kindled analysis in two- and three-point correlation functions of Ly\,$\alpha$ absorbers \citep{soumak_twothreecf_2021}.

\subsection{KODIAQ}
KODIAQ DR2\footnote{\url{https://koa.ipac.caltech.edu/workspace/TMP_939bFW_53591/kodiaq53591.html}} has 300 reduced, continuum-fitted, high-resolution quasar spectra at $0<z<5.3$ with resolving power $R \gtrsim 36\,000$ \citep{lehner_galactic_2014,omeara_first_2015,omeara_second_2017}. The continuum is fitted by hand one echelle order at a time using Legendre polynomials. These high-resolution spectra come in 1.3 or 2.6~km\,s$^{-1}$ velocity spacing. We co-add and resample different observations onto a common 3~km\,s$^{-1}$ grid using exposure times as weight \citep{gaikwad_consistent_2020}.
While this resampling step is not required, it significantly reduces our computational cost while not affecting any of the scales of interest.

\subsection{SQUAD}
SQUAD DR1\footnote{\url{https://doi.org/10.5281/zenodo.1345974}} consists of 467 fully reduced, continuum-fitted high-resolution quasar spectra at redshifts $0<z<5$ with resolving power $R \gtrsim$ 40\,000 \citep{murphy_uves_2019}. The continuum fitting consists of an automatic phase and then a manual phase to eliminate the remaining artefacts. These spectra are sampled onto varying 1.3--3.0~km\,s$^{-1}$ spaced grids in velocity units. As with KODIAQ, we resample these onto a common 3~km\,s$^{-1}$ grid.

There are two further important corrections to the reduced spectra. First, the median seeing is smaller than the slit width for some observations. This results in underestimated nominal resolution values and consequently over-correcting spectrograph resolution. We correct the reported nominal resolution by approximating $R_{\mathrm{cor}}=Rs/\theta$ only when $s>\theta$, where $s$ is the slit width and $\theta$ is the median seeing, both in arc seconds. This yields 25\% correction on average with a maximum of 150\%. The net effect on $P_{\mathrm{1D}}$ is less than 3\% even at $k=0.1$\,s\,km$^{-1}$ since the resolution comes into effect at significantly smaller scales.

Second, \citet{murphy_uves_2019} note that their pipeline underestimates the errors in saturated absorption lines. They provide $\chi^2_\nu$ of each pixel about the weighted mean when combining multiple exposures. Following \citet{king_spatial_2012}'s correction, we apply a median filter of size 5 to $\chi^2_\nu$ and multiply the error with $\sqrt{\mathrm{median}[\chi^2_\nu]}$ if it is greater than 1.

\subsection{XQ-100}
XQ-100 contains 100 quasars at redshifts $3.5<z<4.5$ with resolving power ranging from $R\sim$ 4\,000--7\,000 and spectra from different arms made available \citep{lopez_xq-100_2016}. These spectra are obtained from the ESO database\footnote{\url{http://telbib.eso.org/detail.php?bibcode=2016A\%26A...594A..91L}}. For each arm, the continuum is manually fit by selecting absorption free points. The Ly\,$\alpha$ forest falls into VIS and UVB arms. Spectra from VIS arm have a pixel spacing of 11 km\,s$^{-1}$, whereas UVB spectra are on 20~km\,s$^{-1}$ velocity spaced grids. Given the lower resolution compared with KODIAQ and SQUAD, we do not resample these spectra onto another grid. We also keep the spectra from different arms separate, which allows us to keep resolution correction more accurate. For simplicity, we ignore any correlations between overlapping regions
of these two arms.

Because the seeing is smaller than the slit width for most observations, the nominal resolution is similarly underestimated for this set. We correct the resolution for this effect by interpolating the tabulated values\footnote{\url{https://www.eso.org/sci/facilities/paranal/instruments/xshooter/inst.html}} only when the seeing is smaller than the slit width \citep{yeche_constraints_2017}. We do not extrapolate below the smallest provided slit width value. Even though this yields 30\%  resolution correction on average and doubles the resolution at maximum, the effect in \poned is smaller.
We see an average of 15\% correction for UVB arm and 5\% for VIS arm in \poned at $k=0.045$\,s\,km$^{-1}$, our confidence limit for XQ-100.

\section{Method}
\label{sec:method}
\subsection{Summary of the pipeline}
Before processing the spectra, we need to correct the nominal resolution values, identify duplicate quasar observations and mark DLAs.
We do not mask metal lines, but subtract a statistical estimate of the metal power using side band regions. First, the nominal resolutions of XQ-100 and SQUAD spectra are corrected for seeings that are less than the slit width as described in their respective sections.
Second, assuming a data set does not contain duplicates in itself, we identify quasars within 10$^{\prime\prime}$ of each other from different sets as duplicates, and pick the spectrum that has the highest S/N per km\,s$^{-1}$. 
Third, we visually identify and remove DLAs with the help of a simple automated DLA finder and given catalogs (see Section~\ref{subsec:dla}).

We then process the spectra as follows\footnote{\url{https://bitbucket.org/naimgk/qsotools}. Our pipeline code is not well documented, but we make it public for reference.}:
\begin{enumerate}
    \item Our analysis region is limited to $1.7 < z <4.7$ and 1050~\AA $<\lambda_{\mathrm{RF}}<$ 1180~\AA. We remove pixels that fall outside of these bounds.

    \item These spectra are still susceptible to reduction artefacts such as spikes due to continuum normalization near an echelle order, sky subtraction or cosmic rays. 
    We remove these outlier artifacts by eliminating pixels with flux values one median absolute deviation (MAD) outside 0 and 1, and by eliminating pixels with errors 3.5 MAD above the median error. In other words, we keep pixels that satisfy the following criteria: 
    \begin{align}
        -\mathrm{MAD}(F) &< F(\lambda) < 1+\mathrm{MAD}(F) \\
        0 &< \sigma(\lambda) < \mathrm{median}(\sigma) +  3.5\times \mathrm{MAD}(\sigma),
    \end{align}
    where $\mathrm{MAD}$ is computed in the Ly\,$\alpha$ region. We prefer median statistics because they are robust against outliers.
    
    \item We co-add multiple KODIAQ observations, then resample these and SQUAD spectra onto a common 3 km s$^{-1}$ spaced grid. We keep XQ-100 data in its original spacing, and do not co-add UVB and VIS arms.
    
    \item We divide by the best-fit mean flux of the corresponding data set to get $\delta_F=F/\overline{F} - 1$.

    \item We divide the forest into three equal regions in the rest frame, and split all spectra into these chunks to speed up our calculation and help continuum marginalization. We remove chunks that are shorter than 10\% of the entire forest.
\end{enumerate}

% This process results in 1544 spectral chunks from 552 quasars comprised 183 KODIAQ, 276 SQUAD and 93 XQ-100 quasars when all three sets are combined. Removing XQ-100 from the pool yields 1276 chunks with 186 KODIAQ and 278 SQUAD quasars.
Combining KODIAQ and SQUAD with this process results in 1276 spectral chunks from 464 quasars comprised 186 KODIAQ and 278 SQUAD quasars. We are then left with 74 unique XQ-100 quasars. UVB arm contributes 62 chunks, while VIS contributes 47.

\subsection{Mean flux}
We measure the mean flux of each data set independently and use the respective best fits as mean flux to calculate $\delta_F=F/\overline{F} - 1$. As we discuss in Section~\ref{sec:results}, this removes some systematic errors in continuum fitting procedure.

\begin{figure}
    \centering
    \includegraphics[width=\columnwidth]{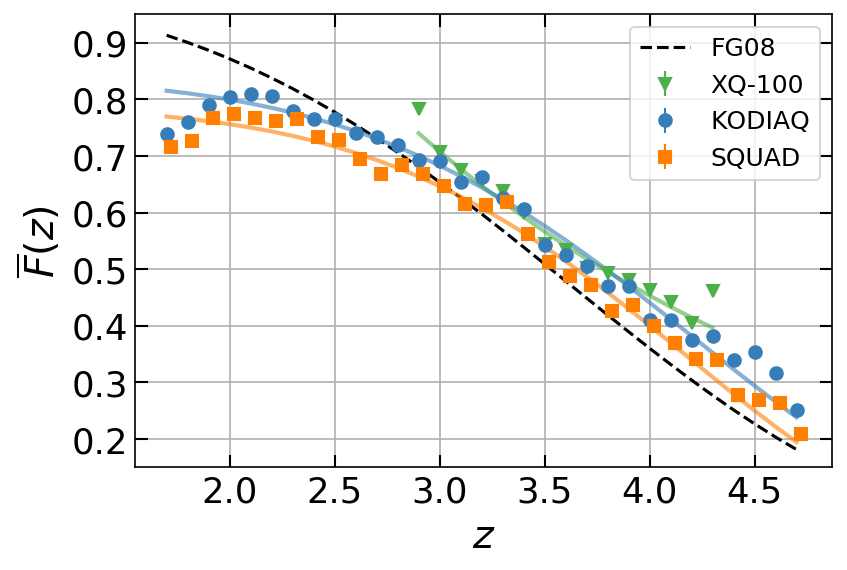}
    \caption{Mean flux measurement of each data set when DLAs are masked. Solid transparent lines show best-fitted results. KODIAQ and SQUAD have deviations at low $z$. \citetalias{faucher-giguere_direct_2008} is extrapolated to lower redshifts. We address the discrepancies in Section~\ref{sec:results}.}
    \label{fig:mean_flux}
\end{figure}

We add Ly\,$\alpha$ variance to the pipeline error on pixel level using a fiducial power from fitting equation~(\ref{eq:pd13_fitting_fn}) to previous measurements and the mean flux from \citet{faucher-giguere_direct_2008} (\citetalias{faucher-giguere_direct_2008}). 
We assign the  square of the mean S/N per km\,s$^{-1}$ of the Ly\,$\alpha$ region as weight, then simply average pixel values. We do not use inverse variance at the pixel level because the pipeline noise and flux are correlated. Specifically, low flux regions almost always have smaller pipeline error estimates. 
Error for a given bin is the propagated value of the modified pipeline errors that go into that bin. 
Note that we do not account for systematics here and therefore underestimate the errors.
We then fit \citet{becker_refined_2013} form for $\overline F(z)=\exp(-\tau(z))$ to our measurements, where $\tau(z)$ is given by
\begin{equation}
    \tau(z) = C + \tau_0\left(\frac{1+z}{1+z_0}\right)^\beta,
\end{equation}
and $z_0=3.5$. Fig.~\ref{fig:mean_flux} shows our mean flux measurements from each data set by using $dz=0.1$ spaced bins.  Table~\ref{tab:mean_flux_fit} has the corresponding best-fitted parameters. 
 
\begin{figure}
    \centering
    \includegraphics[width=\columnwidth]{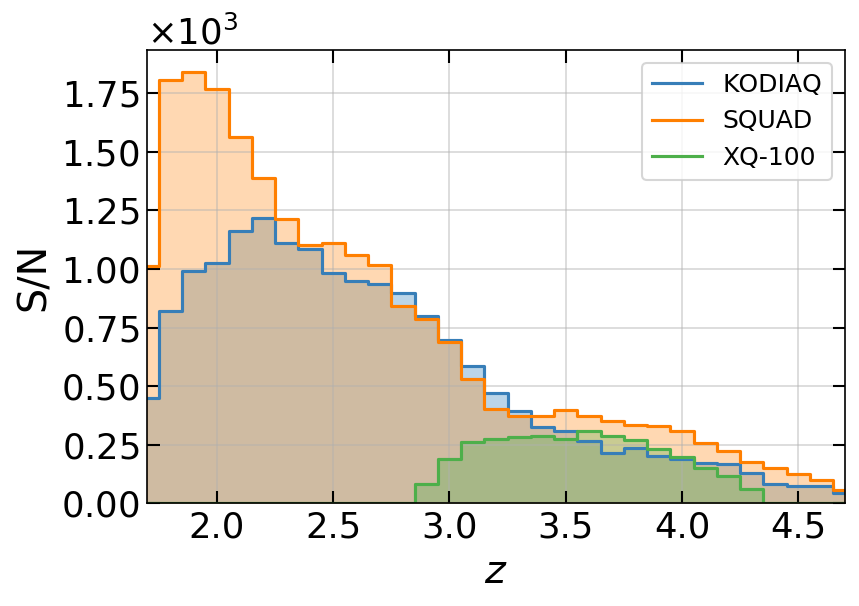}
    \caption{S/N as a function of redshift for each data set, where we define S/N to be $\overline F(z)/\sigma(z)$. The signal mean flux is assumed to be from \citetalias{faucher-giguere_direct_2008}, $\overline F =\overline F_{\mathrm{FG08}}$.
    The noise in a redshift bin $\sigma(z)$ is the propagated error when mean flux is estimated, i.e., the error bars in Fig.~\ref{fig:mean_flux}. Note we do not account for systematics here, and therefore overestimate the S/N.
    }
    \label{fig:stacked_snr}
\end{figure}

Using the errors we obtain from this analysis, we plot the S/N of each data set in Fig.~\ref{fig:stacked_snr}. Since the systematics are not included, we overestimate the S/N, so this figure is for illustration purposes only. We assume signal mean flux to be from \citetalias{faucher-giguere_direct_2008}. XQ-100 equally contributes to S/N distribution between $3\lesssim z \lesssim 4$ even though it has lower resolution.
Additionally, we provide the total redshift path length covered by our final data set in Table~\ref{tab:zpath_length}.

\begin{table}
    \centering
    \begin{tabular}{l|c|c|c}
        \hline
        Set & $\tau_0$ & $\beta$ & $C$ \\
        \hline
        \citetalias{faucher-giguere_direct_2008} & 0.675 & 3.92 & 0 \\
        KODIAQ & 0.373 & 5.13 & 0.18 \\
        SQUAD & 0.377 & 5.54 & 0.24 \\
        XQ-100 & 2.000 & 1.00 & -1.43 \\
        \hline
    \end{tabular}
    \caption{Best-fitted parameters to our mean flux measurements of each data set when DLAs are masked.}
    \label{tab:mean_flux_fit}
\end{table}

\begin{table*}
    \centering
    \begin{tabular}{cl*{15}{c}}
        \hline
\multicolumn{2}{c}{$z$ bin} & 1.8 & 2.0 & 2.2 & 2.4 & 2.6 & 2.8 & 3.0  & 3.2 & 3.4 & 3.6 & 3.8 & 4.0 & 4.2 & 4.4 & 4.6 \\
\hline
Redshift & KS & 14.7 & 18.5 & 18.8 & 16.5 &  18.1 &  17.3 & 14.3 & 7.3 & 5.5 & 5.2 &  5.3 & 5.1 & 3.8 & 1.7 &  1.1\\
 path &XQ-100 & - & - & - & - & - & - & 3.0 & 7.0 & 9.0 & 7.1 & 6.3 & 4.0 & 1.5 & - & - \\
length &Total & 14.7 & 18.5 & 18.8 & 16.5 &  18.1 &  17.3 & 17.3 & 14.3 & 14.5 & 12.3 & 11.6 & 9.1 & 5.3 & 1.7 & 1.1 \\
\hline
    \end{tabular}
    \caption{Total redshift path lengths covered by combined KODIAQ and SQUAD sample (KS) and remaining unique XQ-100 quasars in our analysis $z$ bins.}
    \label{tab:zpath_length}
\end{table*}

\subsection{Quadratic estimator}
Our primary method is the quadratic maximum likelihood estimator (QMLE) \citep{hamilton_towards_1997,tegmark_karhunen-loeve_1997,tegmark_measuring_1998, seljak_weak_1998, mcdonald_ly$upalpha$_2006}. We refer the reader to \citet{karacayli_optimal_2020} for details and extensive tests\footnote{\url{https://bitbucket.org/naimgk/lyspeq}}. We also implemented a simple FFT estimator that does not account for noise or resolution as a rough cross-check for our results.

An important feature of our QMLE implementation is estimating deviations from a fiducial power spectrum such that $P(k, z) = P_{\mathrm{fid}}(k, z) + \sum_{m,n} w_{(mn)}(k, z) \theta_{(mn)}$, where we adopt top-hat $k$ bands with $k_n$ as bin edges and linear interpolation for $z$ bins with $z_m$ as bin centres \citep{font-ribera_how_2018}.  \citet{palanque-delabrouille_one-dimensional_2013} provides a fitting function with best-fitted parameters. We further modify their fitting function with a Lorentzian decay \citep{karacayli_optimal_2020}:
\begin{equation}
    \label{eq:pd13_fitting_fn}\frac{kP(k, z)}{\pi} = A \frac{(k/k_0)^{3 +n + \alpha\ln k/k_0}}{1+(k/k_1)^2} \left(\frac{1+z}{1+z_{0}}\right)^{B + \beta\ln k/k_0},
\end{equation}
where $k_{0} = 0.009$\skm and $z_{0}=3.0$. Combined with \citet{walther_new_2017} results, the best-fitted parameters are $A=0.066, n=-2.685, \alpha=-0.22, B=3.59, \beta=-0.16$ and $k_1=0.053$\,s\,km$^{-1}$. We note that this fit has bad $\chi^2$ and should not be used for scientific purposes, but it is sufficient for a baseline estimate.

Given a collection of pixels representing normalized flux fluctuations $\bm{\delta}_F$, the quadratic estimator is formulated as follows:
\begin{equation}
    \label{eq:theta_it_est}\hat \theta^{(X+1)}_{\alpha} = \sum_{\alpha'} \frac{1}{2} F^{-1}_{\alpha\alpha'}(d_{\alpha'} - b_{\alpha'} - t_{\alpha'}),
\end{equation}
where $X$ is the iteration number and 
\begin{align}
    \label{eq:data_dn} d_{\alpha} &= \bm{\delta}_F^\mathrm{T} \mathbf{C}^{-1}\mathbf{Q}_{{\alpha}} \mathbf{C}^{-1} \bm{\delta}_F, \\
    \label{eq:noise_bn}b_{\alpha} &= \Tr(\mathbf{C}^{-1}\mathbf{Q}_{\alpha} \mathbf{C}^{-1}\mathbf{N}), \\
    \label{eq:signalfid_tn}t_{\alpha} &= \Tr(\mathbf{C}^{-1}\mathbf{Q}_{\alpha} \mathbf{C}^{-1}\mathbf{S}_{\mathrm{fid}}),
\end{align}
where the covariance matrix $\mathbf C \equiv \langle\bm{\delta}_F\bm{\delta}_F^T\rangle$ is the sum of signal and noise as usual, $\mathbf C = \mathbf{S}_{\mathrm{fid}} + \sum_{\alpha} \mathbf{Q}_{\alpha} \theta_{\alpha}+\mathbf{N}$, $\mathbf{Q}_{\alpha} = \partial \mathbf{C} / \partial \theta_{\alpha}$ and the estimated Fisher matrix is
\begin{equation}
    \label{eq:fisher_matrix}F_{\alpha\alpha'} = \frac{1}{2} \Tr(\mathbf{C}^{-1}\mathbf{Q}_{\alpha} \mathbf{C}^{-1} \mathbf{Q}_{\alpha'} ).
\end{equation}
The covariance matrices in the right hand side of equation~(\ref{eq:theta_it_est}) are computed using parameters from the previous iteration $\theta_{\alpha}^{(X)}$. Assuming different quasar spectra are uncorrelated, the Fisher matrix $F_{\alpha\alpha'}$ and the expression in parentheses in equation~(\ref{eq:theta_it_est}) can be computed for each quasar, then accumulated, i.e. $\mathbf{F}=\sum_q\mathbf{F}_{q}$ etc. 

In order to adapt this to Ly$\alpha$ forest analysis, we first convert a pixel's wavelength to velocity using logarithmic spacing as has been the cosmology independent convention,
\begin{align}
    v_i &= c \ln (\lambda_i/\lambda_{\mathrm{Ly\alpha}}) \\
    z_i &= \mathrm{e}^{v_i/c} - 1,
\end{align}
where $\lambda_{\mathrm{Ly\alpha}} = 1216$ \AA.

Second, since the resolution matrix is not provided, we make the approximation that the resolution does not change with wavelength and is Gaussian for the rest of the paper. Then, the signal is the power spectrum multiplied with the spectrograph window function $W(k)$ in Fourier space.
\begin{equation}
    S_{ij}^{\mathrm{fid}} = \int_0^\infty \frac{dk}{\pi} \cos(k v_{ij}) W^2(k) P_{\mathrm{fid}}(k, z_{ij}),
\end{equation}
where $v_{ij}\equiv v_i - v_j$ and $1 + z_{ij} \equiv \sqrt{(1+z_i)(1+z_j)}$. The spectrograph window function is given by
\begin{equation}
    \label{eq:spectrograph_window}W(k) = e^{-k^2R^2/2} \text{sinc}(k\triangle v/2),
\end{equation}
where $R$ is the 1$\sigma$ resolution and $\triangle v$ is the pixel width, both in velocity units. The derivative matrix for redshift bin $m$ and wavenumber bin $n$ is
\begin{equation}
    Q_{ij}^{(mn)} = I_m(z_{ij}) \int_{k_n}^{k_{n+1}} \frac{dk}{\pi} \cos(kv_{ij}) W^2(k),
\end{equation}
where $I_m(z)$ is the interpolation kernel which is 1 when $z=z_m$ and 0 when $z=z_{m\pm1}$. 
We compute these matrices for as many redshift bins as necessary for a given spectrum.

Finally, we assume that the noise of every pixel is independent. This results in a diagonal noise matrix with $N_{ii}=\sigma_i^2$, where $\sigma_i$ is the continuum normalized pipeline noise divided by the mean normalized flux $\overline F(z)$. We use the best-fit mean flux of the corresponding data set for $\overline F(z)$.

\begin{figure}
    \centering
    \includegraphics[width=\columnwidth]{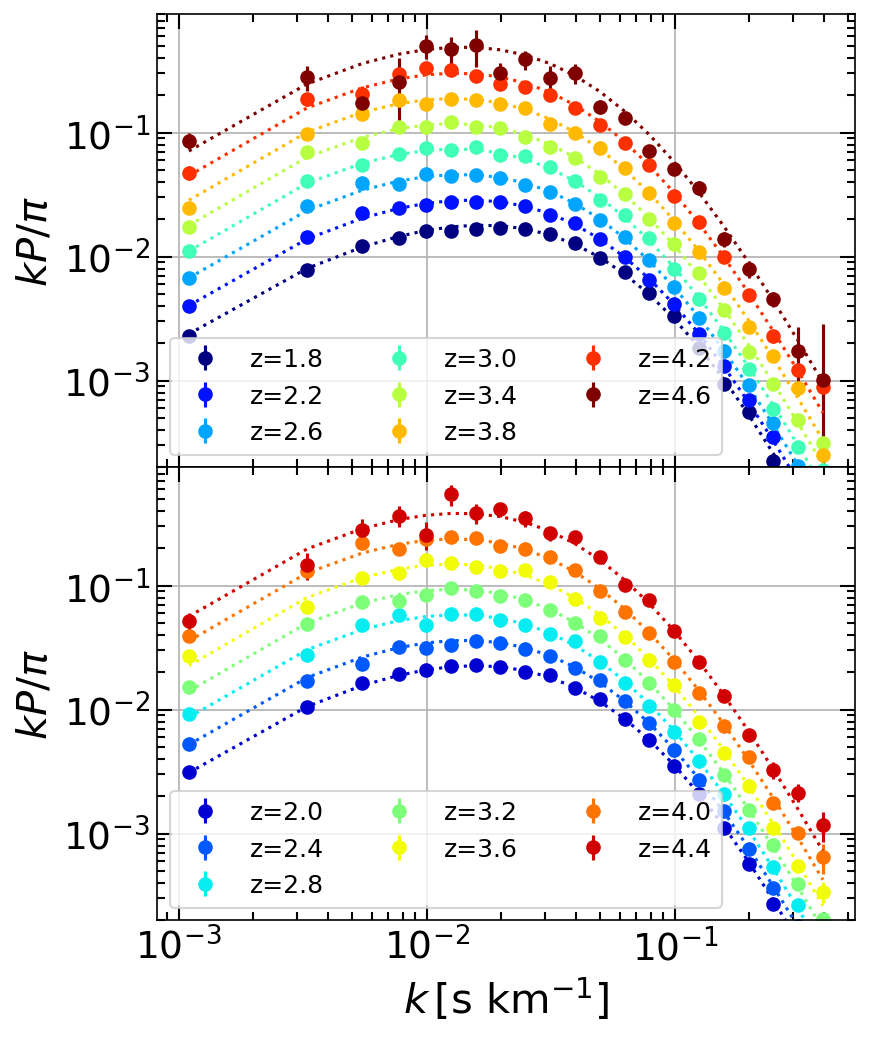}
    \caption{Power spectrum measurement from one mock set (circles) compared to the truth (dotted lines). Error bars are replaced with bootstrap estimates.}
    \label{fig:valid_power}
\end{figure}

\subsection{Validation}
\label{subsec:validation}
Given the resolution and S/N diversity of KODIAQ, SQUAD and XQ-100 data, it is crucial to verify the power spectrum estimates are unbiased and the errors are correctly estimated in an ideal statistical limit.
To validate our method in these both aspects, we generate 100 independent log-normal mock data sets with exact redshift distribution, resolution and noise properties of the data using the procedure described in \citet{karacayli_optimal_2020}. These synthetic spectra approximately produce expected mean flux redshift evolution \citep{faucher-giguere_direct_2008} and power spectrum similar to \citet{palanque-delabrouille_one-dimensional_2013} and \citet{walther_new_2017}. 
Even though these mocks cannot capture all the richness of data, they form the baseline with which we validate QMLE. We use bootstrap error estimates to partially capture effects not present in the mocks. We also discuss some astrophysical and instrumental effects in Section~\ref{subsec:systematics}.

We use the true values for fiducial power spectrum and mean flux to estimate \poned with one iteration\footnote{We note that using a fiducial power that is close to the truth yields better \poned estimates than not using any fiducial \citep{karacayli_optimal_2020}.}. Fig.~\ref{fig:valid_power} shows a sample result from one mock set. From the average of these 100 estimates, we find that our results are unbiased in the range of interest $k<0.1$\skm, but they diverge from the truth for $k\gtrsim 0.2$\skm. We note that these scales are noise-dominated, and further complicated by resolution effects and narrow metal lines in real data. Therefore, they are significantly hard to measure and have been absent in previous studies.
% Furthermore, our bins end at $k\sim 0.4$\skm, whereas these spectra have power until $k_{\mathrm{Ny}} \sim 1$\skm. We suspect this extra, small-scale power leaks into $0.2$\skm $< k < 0.4$\skm range.

\begin{figure}
    \centering
    \includegraphics[width=\columnwidth]{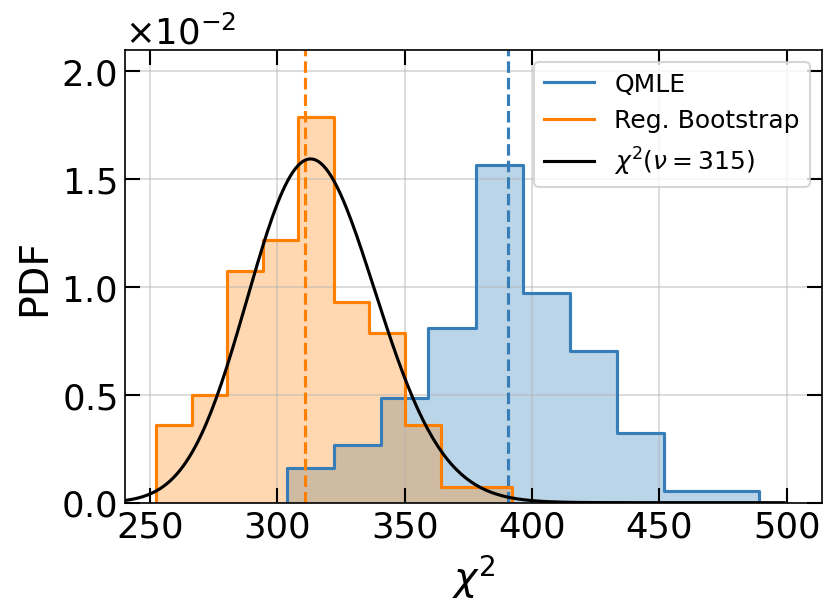}
    \caption{To validate our method, we performed $\chi^2$ analysis on 100 mock sets using the Fisher matrices from QMLE and 25\,000 bootstrap realizations from one set. We regularized the bootstrap Fisher matrix by exploiting its sparsity pattern and by flooring the eigenvalues to their Gaussian limits, which is noted as "Reg. Bootstrap". This produces the expected $\chi^2$ distribution.}
    \label{fig:valid_chisquare}
\end{figure}

We then investigate the accuracy of estimated errors by a $\chi^2$ analysis, and set the average of 100 results as the truth. First, we calculate the $\chi^2$ of each result using the Fisher matrix from QMLE\footnote{We note that for all sets QMLE gives the same Fisher matrix, and so the same covariance matrix.}, which yields larger values than expected. We identify the cause to be the violation of QMLE Gaussianity assumption at small scales by generating Gaussian mocks, which yields the correct $\chi^2$ distribution.
Therefore, we get another estimate of the covariance matrix by generating 25\,000 bootstrap realizations from one set. We find that the diagonals of the bootstrap covariance matrix converges rapidly, but the off-diagonal terms remain noisy (and probably correlated) even with 250\,000 realizations. We believe our sample size of approximately 1500 chunks is not enough for the degrees of freedom in the covariance matrix, which is a 315x315 matrix, or for its condition number $\mathcal{O}(10^{11})$. To achieve a stable covariance matrix, we apply a two-step regularization scheme.

After generating bootstrap realizations with spectral chunks, our final algorithm is based only on data and as follows:
\begin{enumerate}
    \item To prevent off-diagonal noise from leaking, we directly estimate the Fisher matrix element-wise, where $\left|r_{ij}^{\mathrm{QMLE}}\right|>0.01$ and $r_{ij}\equiv F_{ij}/\sqrt{F_{ii}F_{jj}}$ using the algorithm in \citet{padmanabhan_estsparse_2016}. This algorithm further refines the element-wise estimate to the "closest" positive-definite matrix.
    
    \item\label{step:eig_reg} We then find the eigenvalues $\lambda_i$ and eigenvectors $\bm e_i$ of this Fisher matrix. We calculate the precision of these eigenvectors under Gaussianity: $\lambda_i^{\mathrm{QMLE}} = \bm e_i^T \mathbf{F}^{\mathrm{QMLE}} \bm e_i$. This is the theoretical maximum (minimum for the covariance), so we replace $\lambda_i\rightarrow \mathrm{min}(\lambda_i, \lambda_i^{\mathrm{QMLE}})$ and rebuild the Fisher matrix \citep{mcdonald_ly$upalpha$_2006}.
\end{enumerate}

We note that using the bootstrap eigenvectors as basis in step~\ref{step:eig_reg} captures the non-Gaussian mode couplings. However, bootstraps are also missing fluctuations in certain modes due to low statistics; for example, there are only 5 chunks in the last redshift bin. This eigenvalue regularization scheme takes these modes to their Gaussian limit, and is a slight modification of \citet{mcdonald_ly$upalpha$_2006}.

Fig.~\ref{fig:valid_chisquare} shows that the regularized bootstrap Fisher matrix produces the expected $\chi^2$ distribution. These tests give us the confidence that our power spectrum estimates are unbiased, and the covariance can be estimated using bootstrap method at the scales of interest.

We tested our regularization scheme on the bootstrap covariance matrix and observed a marginally worse performance. The element-wise estimate is not positive-definite, but eigenvalue regularization fixes that problem. However, the Fisher matrix has a simpler structure than the covariance matrix. Furthermore, the $\chi^2$ calculation is a linear operation with the Fisher matrix elements which further justifies the element-wise approximation, whereas inverting the covariance matrix is highly non-linear. Therefore, we opted for a direct estimation of the Fisher matrix in this paper.

\begin{figure}
    \centering
    \includegraphics[width=\columnwidth]{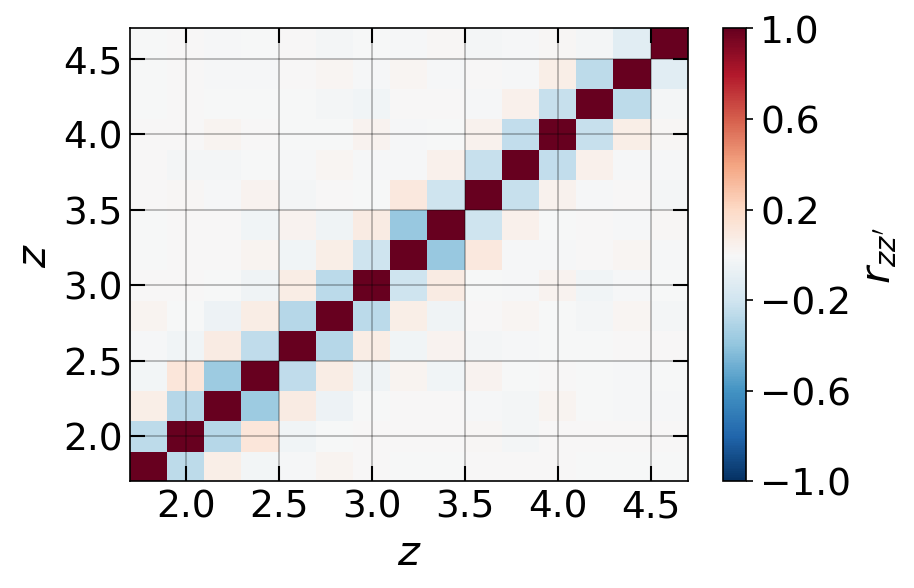} 
    \caption{The normalized bootstrap covariance matrix of data for $k=0.013$\,s\,km$^{-1}$ bin. Neighbouring redshift bins are anti-correlated because of assigning pixel pairs into two redshift bins.}
    \label{fig:lya_boot_cov}
\end{figure}

We also test our mean flux measurement process with one mock set. We find that we obtain the correct mean flux, but the propagated errors are notably underestimated even with added Ly\,$\alpha$ variance, which can be suspected from Fig.~\ref{fig:mean_flux} as well. We remind the reader that we are not interested in a precise mean flux measurement, but only in using it to remove some continuum fitting systematics from spectra. However, we perform another test to quantify the underestimation of these errors. We resample the spectra onto a coarse grid (300 km\,s$^{-1}$) using inverse variance (note signal and errors are not correlated for the mocks). We then estimate the standard deviation using these coarse pixels in a given redshift bin, and find that they are four times the propagated errors. We also note that the correlations between pixels actually matter for the resolution and S/N of our data when resampling  \citep{slosar_measurement_2013}, but it is still unclear if they could yield correct error bars when propagated. 
We speculate that the systematics are most likely a significant source of error for the measurement from data however.

\begin{figure*}
    \centering
    \includegraphics[width=0.95\linewidth]{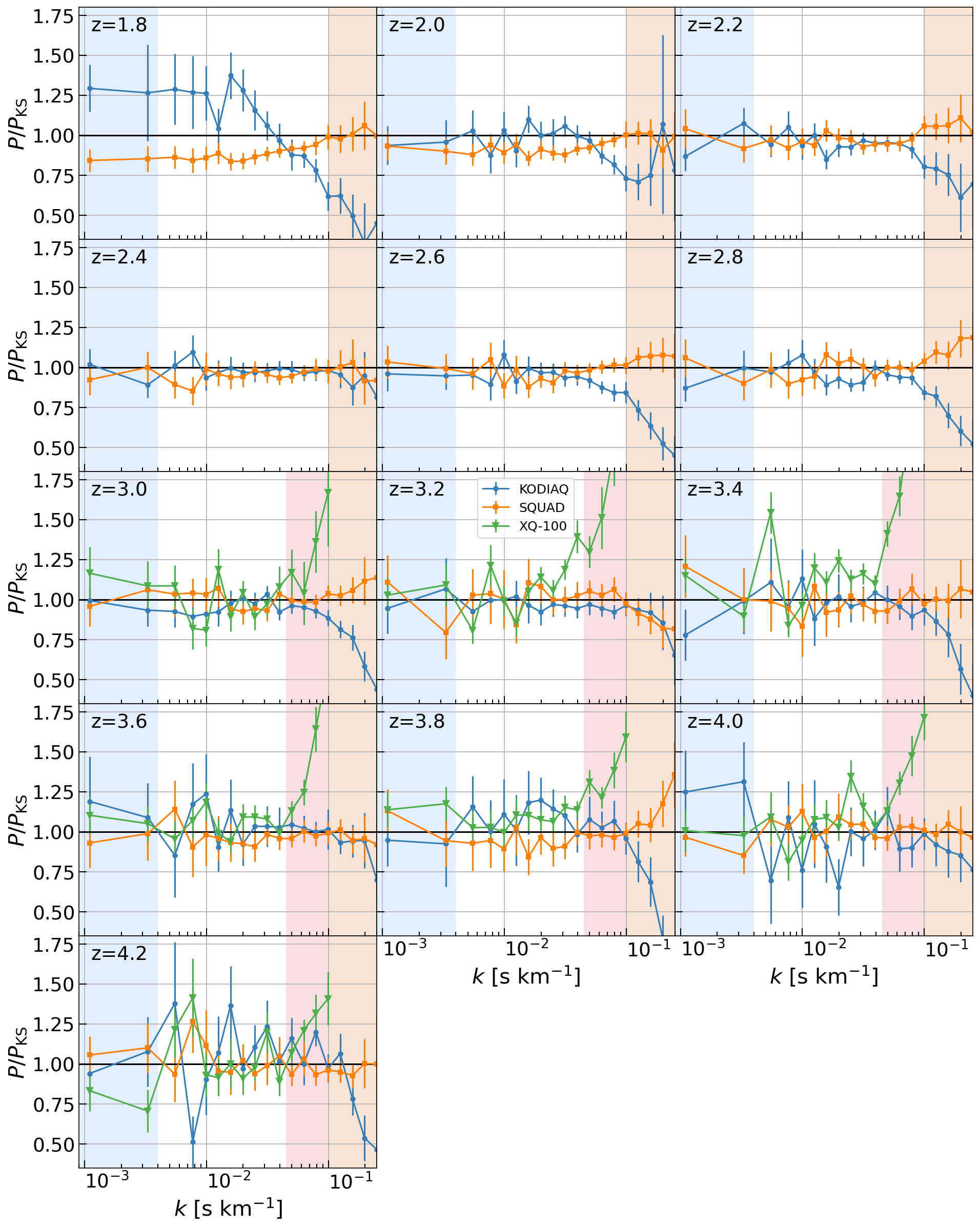}
    \caption{The raw power spectrum measurements of KODIAQ, SQUAD and XQ-100 divided by the raw power from KODIAQ + SQUAD ($P_\mathrm{KS}$) when DLAs are masked. Error bars are from bootstrap. We highlight the large-scale modes $k \lesssim 4\times 10^{-3}$\,s\,km$^{-1}$ that are susceptible to continuum errors in light blue, and small-scale modes $k\gtrsim 0.1$\,s\,km$^{-1}$ that are affected by noise, metal contamination and resolution effects in light orange. Light pink region is excluded from our XQ-100 analysis.}
    %in \citet{irsic_lyman_2017}.}
    %and \citet{yeche_constraints_2017}.}
    \label{fig:ind-runs-compare}
\end{figure*}

% \begin{figure*}
%     \centering
%     \includegraphics[width=0.95\linewidth]{plots/compare-combined-lya-ratio-grid-boot-err-inxqonly.png}
%     \caption{Comparing the effect of adding XQ-100 quasars to the combined data set as the ratio of raw power spectrum measurements when DLAs are masked. XQ-100 is in great tension with KODIAQ + SQUAD (KS) measurement; and biases the final results when included. Error bars are from bootstrap for KS.}
%     \label{fig:combined-compare}
% \end{figure*}

\begin{figure*}
    \centering
    \includegraphics[width=0.95\linewidth]{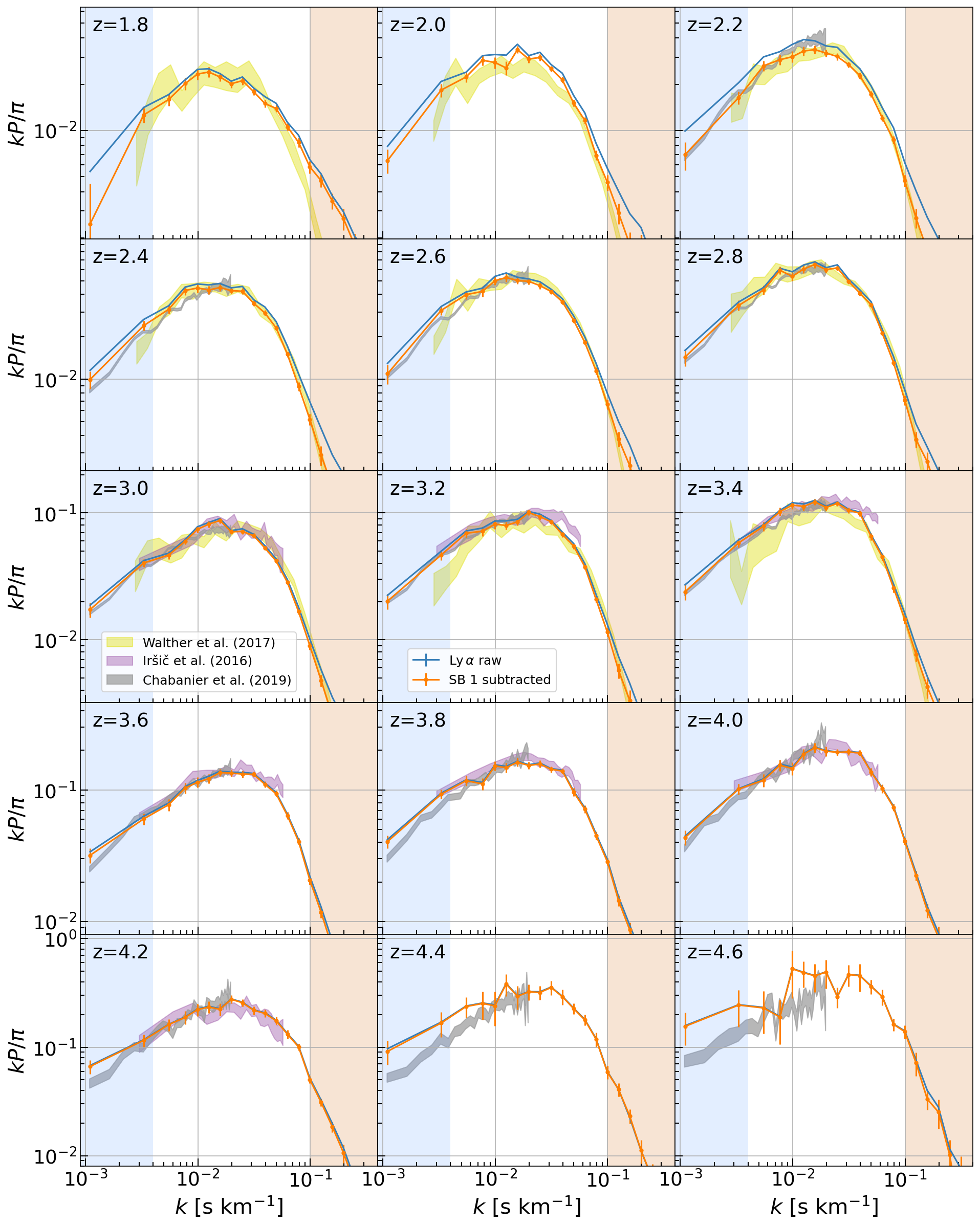}
    \caption{Our final raw and side band subtracted power spectrum measurements when DLAs are masked, where all three sets are combined as described in Section~\ref{subsec:final-res}. The error bars are from bootstraps; and systematic error budget is added in quadrature.
    % Extrapolated fiducial SB power is subtracted from $z=1.8$. 
    We highlight the large-scale modes $k \lesssim 4\times 10^{-3}$\,s\,km$^{-1}$ that are susceptible to continuum errors in light blue, and small-scale modes $k\gtrsim 0.1$\,s\,km$^{-1}$ that are affected by noise, metal contamination and resolution effects in light orange. We remind the reader that the last two $z$ bins have low statistics as well as further continuum fitting complications.}
    \label{fig:p1d_dla_mask}
\end{figure*}

\section{Results}
\label{sec:results}
We measure the power spectrum in 15 redshift bins from $z=[1.8, 4.6]$ with $dz=0.2$ spacing for 21 $k$ bins of which the first 4 are linearly spaced with $\Delta k_\mathrm{lin}=0.0022~$s\,km$^{-1}$ and the rest are logarithmically spaced with $\Delta k_\mathrm{log}=0.1$. 

To remind the reader our pipeline, we mask DLAs by using the accompanying catalogs and visually identifying remaining absorbers with damping wings. We do not mask metal lines, but provide a statistical estimate of the metal contamination using side band regions. Furthermore, we marginalize out the constant and the slope ($\ln \lambda$) terms of the continuum errors. Because the continuum is fitted piece-wise and corrected manually, these are not the exact expressions for the continuum errors, but the marginalization should still remove some or most of the contamination in the largest scales. These points and their contribution to systematic error budget are further discussed below.

We perform an initial run on the combined set with the fiducial power spectrum as described in Section~\ref{sec:method}. We fit the results for $k<0.1$\,s\,km$^{-1}$ to get a new estimate for the fiducial power, and then perform one iteration. The new parameters are $A=0.084, n=-2.655, \alpha=-0.155, B=3.64, \beta=0.32$ and $k_1=0.048$\,s\,km$^{-1}$. We showed this process yields better power spectrum estimates in \citet{karacayli_optimal_2020}.

The estimated errors from QMLE are under Gaussian assumption and significantly underestimated at small scales and high redshifts. 
We estimate the error bars by generating bootstrap realizations and using our regularization scheme as discussed in Section~\ref{subsec:validation}. 
Since our method divides pixel pairs into two redshift bins, the power spectrum estimates are correlated between these redshift bins. 
Fig.~\ref{fig:lya_boot_cov} shows an example normalized covariance matrix from bootstrap estimation for $k=1.3\times 10^{-2}$\,s\,km$^{-1}$ bin.

\subsection{Consistency between data sets}
Our aim in this work is to utilize the combined data for the best S/N, but first we compare individual power spectrum estimates to check our treatment of the data and consistency between data sets.
Fig.~\ref{fig:ind-runs-compare} lays out the ratios of individual power spectra to our raw measurement from KODIAQ + SQUAD (KS).

The first notable difference is between KODIAQ and SQUAD at $z=1.8$, but they both still fall within the large errors of \citet{walther_new_2017}. 
% This bin will actually be outside our conservative results since it does not have a metal power estimate.
Otherwise, these two sets agree with each other.
To quantify the agreement between them, we calculate $\chi^2$ from the power difference:
\begin{equation}
    \chi^2 = \bm{d}^T (\mathbf{C}_1 +\mathbf{C}_2)^{-1} \bm{d},
\end{equation}
where $\bm{d} = \bm P_1 - \bm P_2$.
This yields $\chi^2 = 218$ for 180 degrees of freedom (dof) while removing $z=1.8$ and $k>0.1$\skm bins. This value is $2\sigma$ away from the mean; or in other words, probability of getting $\chi^2>218$ is 3\%.
This is a reasonable probability when comparing two data sets, so we conclude they are consistent and remove $z=1.8$ from our conservative range\footnote{We have a number of hypotheses for this discrepancy. First, the Ly\,$\alpha$ forest at low redshifts will be in the UV where the response of most spectrographs rapidly falls. Also, metal contamination would have the largest effect at these redshifts and is hard to measure for these samples. However, we were not able to isolate a particular cause, which is why we exclude this redshift bin from our final results.}.

We find that our XQ-100 results agree with \citet{irsic_lyman_2017} in their provided range.
The largest inconsistency here most visibly comes from $k>0.06$\skm values.
We remind the reader that we corrected the resolution for seeing to the best of our knowledge, but it can still carry some errors especially at this range given the Nyquist frequency for XQ-100 is $k\sim0.1$\skm. 
So, accurately measuring these modes is an ambitious goal for XQ-100 alone.
Given our limited understanding of the resolution, we assign a conservative $k<0.045$\skm range for our XQ-100 measurement, where $\chi^2=83$ for 77 dof when systematic errors are included.
Unfortunately, this makes combining XQ-100 at the chunk level complicated as QMLE in its current form cannot account for systematic errors.
We defer how these systematic errors can be incorporated into QMLE to a future work.
Instead, we provide a separate measurement and a covariance matrix for XQ-100 in this paper.

% However, most importantly we find that XQ-100 is in tension with the other two data sets.
% The $\chi^2$ between XQ-100 and KS results for $k<0.06$\skm is 153 for 84 dof.
% The probability of getting $\chi^2>153$ is 0.0006\%. 

% Our most important conclusion in this section is that XQ-100 is in significant tension with KS; and biases the final results when included.
% Therefore, we exclude it from our final measurement.

\subsection{Final combined results}
\label{subsec:final-res}
% We combine KS and XQ-100 measurements with a weighted average. 
Let us start this section with a review of the weighted average and its error propagation. Assume for a data set $m$, we have measurements $\bm x_m$ with respective errors $\mathbf{C}_m$. Using normalized weights $\mathbf{W}_m$, the weighted average $\bm x_W$ and resulting error $\mathbf{C}_W$ are given by
\begin{align}
    \bm x_W &= \sum_m \mathbf{W}_m \bm x_m, \\
    \label{eq:cov_weighted}\mathbf{C}_W &= \sum_m \mathbf{W}_m \mathbf{C}_m \mathbf{W}_m^T + \sum_{m \neq n} \mathbf{W}_m \mathbf{C}_{mn} \mathbf{W}_n^T,
\end{align}
where $\mathbf{C}_{mn}$ is the cross-covariance between measurements $m$ and $n$.
These become the usual inverse variance weighted average expressions with $\mathbf{W}_m \propto (\mathbf{C}_m)^{-1}$ if the measurements are uncorrelated.
In our case however, some systematic errors (DLAs, continuum and metals) between KS and XQ-100 are data-independent and correlated, whereas resolution systematics for example are not.
Let us define $\mathbf{\Lambda}^\mathrm{cor}$ to be these data-independent, correlated systematics between KS and XQ-100, and define $\mathbf{\Lambda}^\mathrm{uncor}_m$ to be the data-dependent, uncorrelated systematics for measurement $m$.
Then, we substitute $\mathbf{C}_{mn}= \mathbf{\Lambda}^\mathrm{cor}$ and $\mathbf{C}_{m}=\mathbf{C}_m^\mathrm{stat}+\mathbf{\Lambda}^\mathrm{uncor}_m + \mathbf{\Lambda}^\mathrm{cor}$, and correlated systematic comes out since $\sum_m \mathbf{W}_m \equiv \mathbf{1}$.
\begin{equation}
    \mathbf{C}_W = \sum_m \mathbf{W}_m (\mathbf{C}_m^\mathrm{stat}+\mathbf{\Lambda}^\mathrm{uncor}_m) \mathbf{W}_m^T + \mathbf{\Lambda}^\mathrm{cor}
\end{equation}
In our case, $\mathbf{\Lambda}^\mathrm{cor} = \mathbf{\Lambda}^{\mathrm{DLA}} + \mathbf{\Lambda}^{\mathrm{Cont}} + \mathbf{\Lambda}^{\mathrm{Metal}}$ and $\mathbf{\Lambda}^\mathrm{uncor}_m$ is the resolution systematics.
In short, uncorrelated systematics will become smaller when different measurements are combined, while correlated systematics stay the same.

We now use the inverse variance weights with systematic errors, which also prevents underestimated statistical errors from dominating the average:
\begin{equation}
    \mathbf{W}_m \propto (\mathbf{C}_m^\mathrm{stat}+\mathbf{\Lambda}^\mathrm{uncor}_m + \mathbf{\Lambda}^\mathrm{cor})^{-1}
\end{equation}
We again generate 25\,000 bootstrap realizations to calculate the statistical Fisher matrix. 
Our systematic error budget is the same for KS and XQ-100 measurements except for the resolution (see Section~\ref{subsec:systematics}) and is added in quadrature to the diagonal. 
We also add large numbers to the diagonal of XQ-100 covariance at $k>0.045$\skm to remove these modes from the average. 
For both measurements, we also subtract the metal power and add its covariance first, so that the weights have all the statistical and systematic errors in place.

Our results from this weighted average are shown in Fig.~\ref{fig:p1d_dla_mask}, which is in good agreement with previous measurements even at the largest scales \citep{irsic_lyman_2017, walther_new_2017, chabanier_one-dimensional_2019}. Note that we do not directly compare with \citet{gaikwad_consistent_2020} since their method involves forward modelling the spectra and does not subtract noise nor deconvolve the spectrograph window function. We caution the reader that the number of quasars and different continuum treatments limit the accuracy of our measurement for $k \lesssim 4\times 10^{-3}$\,s\,km$^{-1}$; and refer the reader to eBOSS \citep{chabanier_one-dimensional_2019} for a better measurement at these scales. Nevertheless, we are encouraged by this agreement at the large scales.

\begin{figure}
    \centering
    \includegraphics[width=\columnwidth]{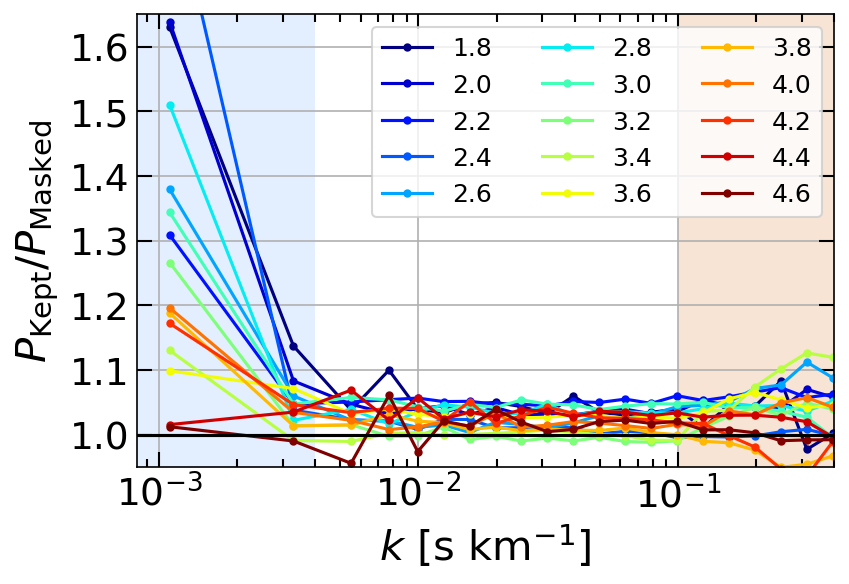}
    \caption{The ratio of the measured power spectrum from the combined data set between DLAs kept $P_{\mathrm{Kept}}$ and DLAs masked $P_{\mathrm{Masked}}$. The damping wings add power to large scales, and the lower mean flux overcomes the full absorption suppression from medium to small scales by changing it $\approx 5\%$.}
    \label{fig:dla_over_nodla}
\end{figure}

\subsection{Damped Ly\,\texorpdfstring{$\alpha$}{Lg} absorbers}
\label{subsec:dla}
The net result of damped Ly\,$\alpha$ absorbers (DLA) on the power spectrum comprises three effects. The primary effect is the large amount of power added to the largest scales due to their damping wings. The central region of complete absorption has two competing effects: suppression of power and amplification because of lower mean flux measurement \citep{mcdonald_physical_2005, rogers_simulating_2018}. These high-column density absorbers are difficult to model and simulate, so they are often removed from spectra. We describe our DLA identification and removal in this section.

The SQUAD data set comes with a DLA catalog; and the ones in XQ-100 are identified in \citet{sanchez-ramirez_evolution_2016}. We mask a reported DLA at $z_\mathrm{abs}$ between $[\lambda_C -W/2, \lambda_C +W/2]$, where $\lambda_C=(1+z_\mathrm{abs})\lambda_{\mathrm{Ly\alpha}}$ and $W$ is the equivalent width given by the following equation \citep[Sec. 16.4.4]{mo_galaxy_2010}:
\begin{equation}
    W = 7.3 (1+z_\mathrm{abs}) \sqrt{\frac{N_\mathrm{H\textsc{i}}}{10^{20}~\mathrm{cm}^{-2}}} \text{~\AA} \label{eq:equiv_width}
\end{equation}

We also apply a simple automated DLA finder which first finds regions that are consecutively $F(\lambda) < \overline F(\lambda) + \sigma(\lambda)$. If a region is longer than the equivalent width for an absorber with $N_\mathrm{H\textsc{i}}=10^{19}~$cm$^{-2}$ in equation~(\ref{eq:equiv_width}) at the central redshift, we mark it as a DLA candidate. We then visually inspect all these candidates for damping wings to remove from data. 

Our results in Fig.~\ref{fig:p1d_dla_mask} are already without DLAs. We performed a run where we kept in the DLAs as well. Fig.~\ref{fig:dla_over_nodla} shows the effect described in the beginning of this section. The presence of DLAs significantly affects the largest scales $k\lesssim 3\times 10^{-1}$\skm, while boosting the intermediate scales by few percent because of lower mean flux estimates.

\begin{figure}
    \centering
    \includegraphics[width=\columnwidth]{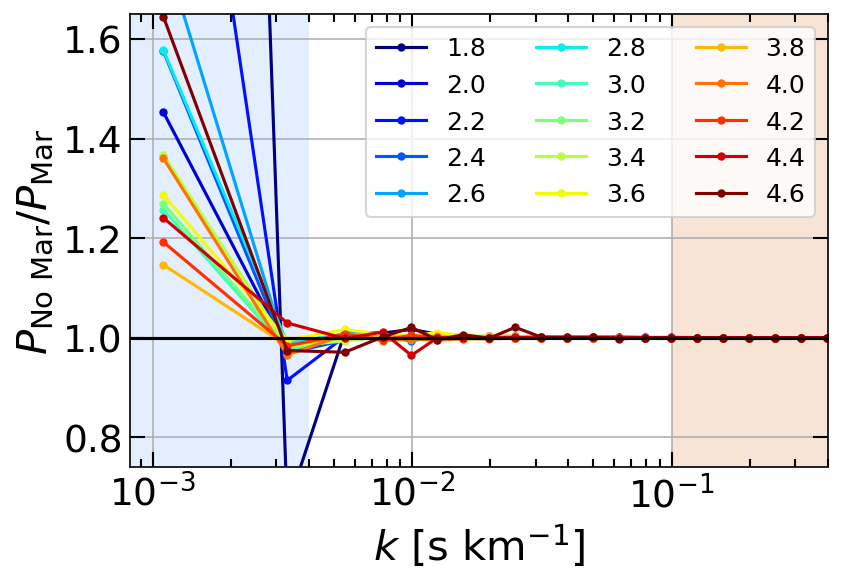}
    \caption{The ratio of the measured power spectrum without continuum marginalization $P_{\mathrm{No\,\,Mar}}$ to with marginalization $P_{\mathrm{Mar}}$. Continuum errors significantly contaminate the largest scales; the rest fluctuates around less than one percent. }
    \label{fig:cont_marg_ratio}
\end{figure}

\subsection{Continuum}
The observed flux $f(\lambda)$ is divided by the quasar continuum $C(\lambda)$ and the mean normalized flux $\overline F(\lambda)$ to obtain the flux fluctuations $\delta_F(\lambda)$.
Since $C(\lambda)$ is smooth,  errors in this process propagate to mostly large scales and are called the continuum errors. 

We start our discussion by considering a systematic scaling of the continuum in a given data set. In other words, we assume the fitted continuum of a quasar $q$ is the true continuum times a constant, $C^*_q(\lambda)=aC_q(\lambda)$, for all quasars in the set. This inevitably scales the normalized flux by the same factor, $F_q^*=f_q/C_q^* = F_q/a$, and hence the measured mean flux $\overline{F}^*=\overline{F}/a$ since $a$ does not depend on $q$. Therefore, using the measured mean flux $\overline{F}^*$ of a given data set removes this systematic bias from the fluctuations $\delta_{F,q}^*=F^*_q/\overline{F}^*-1=\delta_{F,q}$. In fact, we see this type of measured mean flux scaling across data sets, where KODIAQ has 6\% and XQ-100 has 15\% excess mean flux on average compared to SQUAD. We are further encouraged by the agreement of $P_{\mathrm{1D}}$ without any scaling between data sets. Finally, note that this multiplicative scaling would cancel even if it depended on observed wavelength $a=a(\lambda)$ as long as $a$ does not change from quasar to quasar.
%These figures are provided in the supplemental material.

Now, let us introduce a quasar dependent error $\eta_q(\lambda)$ that remains after this systematic scaling, i.e. $C^*_q(\lambda)=aC_q(\lambda)[1+\eta_q(\lambda)]$. Since the continuum itself is smooth, $\eta_q(\lambda)$ will also be smooth and slowly changing given a well-behaving continuum fitting procedure. The effect of this quasar dependent error will propagate to the mean flux measurement by some average.
\begin{align}
    \overline{F}^* &= \frac{\overline{F}}{a} \left\langle \frac{1}{1+\eta_q}\right\rangle_q \approx \frac{\overline{F}}{a} (1- \bar\eta),
\end{align}
where we have assumed $\eta_q$ is small and not correlated with $F_q$. Then,
\begin{align}
    \delta_{F,q}^* &= \frac{F_q^*}{\overline{F}^*} - 1 = \frac{F_q/\overline{F}}{(1+\eta_q)(1- \bar\eta)} - 1 \\
    &= (1+\delta_{F,q})(1-\Delta\eta_q)-1.
\end{align}
Finally,
\begin{equation}
    \delta_{F,q}^*(\lambda) = \delta_{F,q}(\lambda) [1-\Delta\eta_q(\lambda)] - \Delta\eta_q(\lambda),
\end{equation}
where we have defined $\Delta\eta_q(\lambda)\equiv \eta_q(\lambda)-\bar\eta(\lambda)$. Unfortunately, this error does not fully cancel, but it is possible to partially marginalize out the additive term $\Delta\eta_q(\lambda)$ by approximating its form. This error is described by a constant shift and a slope in eBOSS pipeline, $\Delta\eta_q(\lambda)=a_q+b_q\ln \lambda$. We marginalize out these modes in this work as well, even though they are not the exact expressions for the continuum errors because the continuum is fitted piece-wise and then corrected manually. Dividing the forest into three chunks also helps by limiting the wavelength range of $\Delta\eta_q(\lambda)$  that needs to be described. Note that since $\Delta\eta_q(\lambda)$ is smooth, this error affects mostly large scales even when uncorrected. We ignore the multiplicative term within our approximation. Its major complication would stem from the correlations between $\eta_q(\lambda)-F_q(\lambda)$ and $\eta_q(\lambda)-\eta_q(\lambda')$, which can ultimately be treated by smaller errors or by more uniform continuum fitting procedures.

We performed another run where we turned off the continuum marginalization to show its effect. Fig.~\ref{fig:cont_marg_ratio} shows the largest scales $k\lesssim 3\times 10^{-3}$\skm are again significantly affected by the continuum. However, the intermediate scales deviate by less than one percent.

\begin{figure}
    \centering
    \includegraphics[width=\columnwidth]{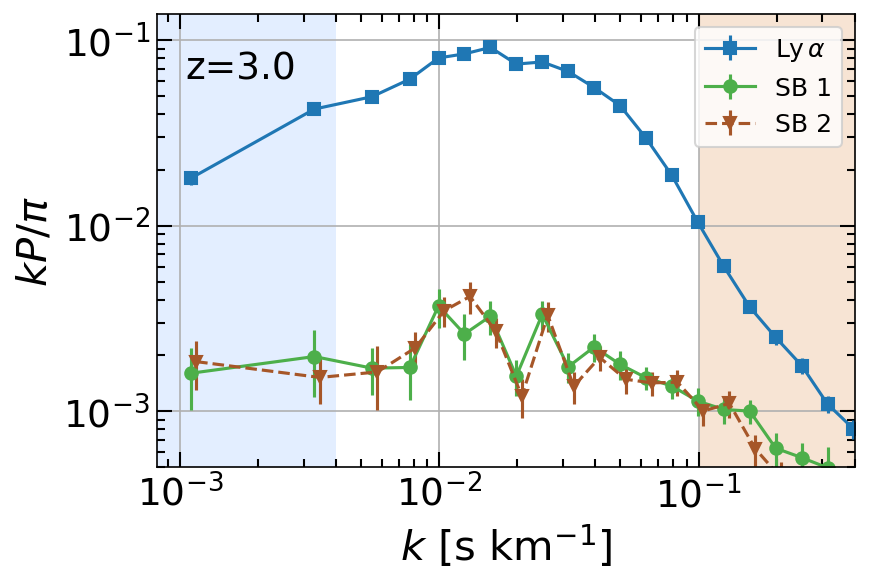}
    \caption{The power spectra measured from the two side bands at $z=3.0$. The C\,\textsc{iv} doublet causes the oscillation peaks at $k=1.3\times 10^{-2}$\,s\,km$^{-1}$ and $2.5\times 10^{-2}$\,s\,km$^{-1}$. This doublet and others are apparent in the correlation function (see Fig.~\ref{fig:xi1d_sbboth_doublets}).}
    \label{fig:SB_power}
\end{figure}

\subsection{Side band power}
The red side of the Ly\,$\alpha$ line in the spectrum is free from H\,\textsc{i} absorption, so this region is used to estimate the power from metals and other systematics \citep{mcdonald_ly$upalpha$_2006, palanque-delabrouille_one-dimensional_2013, chabanier_one-dimensional_2019}. We define the first side band (SB 1) region to be between 1268 -- 1380~\AA, below the Si\,\textsc{iv} line, in quasar's rest frame, and the second side band region to be between 1409 -- 1524~\AA, below the C\,\textsc{iv} line. We use $\overline{F}(z)=1$, and estimate the power on the same $k$ bins.
% starting from $z=2.0$ for SB 1 and $z=2.2$ for SB 2. 
We note that this only removes power due to metals with $\lambda_{\mathrm{RF}} \gtrsim 1400$\,\AA, and hence some metal contamination still remains and produces oscillatory features such as Si~\textsc{iii}-Ly\,$\alpha$ cross correlation \citep{mcdonald_ly$upalpha$_2006, palanque-delabrouille_one-dimensional_2013}.

We limit the side band power estimate to KODIAQ and SQUAD data sets to not further complicate analysis. 
The final metal power is estimated from the Si\,\textsc{iv} region (SB 1), and uses 1391 chunks with 221 KODIAQ and 271 SQUAD quasars.

The initial runs for both side band regions use 10\% of the fiducial power in Section~\ref{sec:method}. We then switch to the best-fitted parameters for $k<0.1$\,s\,km$^{-1}$ as the new fiducial and estimate side band powers with one iteration. 
These new parameters are $A=0.0027, n=-2.92, \alpha=-0.174, B=0.236, \beta=-0.01$ where the Lorentzian term can be ignored as $k_1\sim 10^{10}$\,s\,km$^{-1}$.
We again estimate the covariance matrix using 25\,000 bootstrap realizations and use these instead of QMLE errors. 
We find that the two side band powers mostly agree with each other with some offset at $z=2.0, 3.2$ and 3.6 bins, while $z=2.2$ has a larger offset.
Fig.~\ref{fig:SB_power} shows and compares the side band powers to the Ly\,$\alpha$ power at $z=3.0$. 

Noteworthy features in both side band powers are the two visible peaks due to oscillations in all redshift bins at $k=1.3\times 10^{-2}$\,s\,km$^{-1}$ and $2.5\times 10^{-2}$\,s\,km$^{-1}$. While bootstrap method shows that the QMLE errors are underestimated for $z<3$ and overestimated for $z>3$, these peaks remain visible in QMLE error to bootstrap error ratio. We find these oscillations are due to C\,\textsc{iv} doublet, which is manifested as a peak in the 1D correlation function at $v\approx 500$\kms separation. Moreover, we identified more doublet features in the 1D correlation function with a simple model. 
First, we estimate the correlation function by inverse FFTing the power spectrum before binning into $k$ bins. This yields the correlation function on the full-resolution grid, which we then bin into $v$ bins. Doublets of the same absorber manifest as peaks in this correlation function. To increase S/N, we average all redshift bins. 

\begin{figure}
    \centering
    \includegraphics[width=\columnwidth]{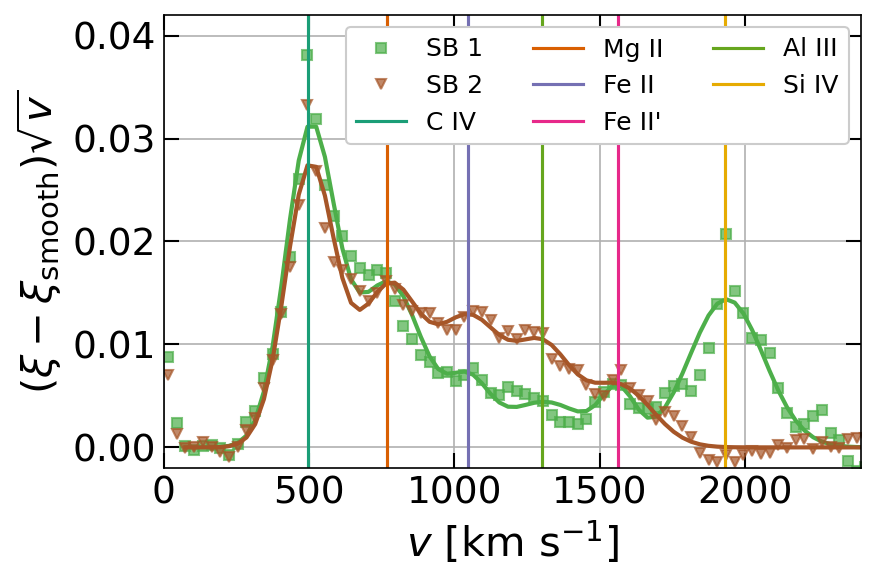} 
    \caption{The residual correlation functions for SB 1 (green squares) and SB 2 (brown triangles). Curves with solid lines are our fitting model for known metal doublets. Our simple fitting function captures the important features. Note again that Si~\textsc{iv} line is outside of SB 2, so there is no peak at $v\approx 2000$\kms.}
    \label{fig:xi1d_sbboth_doublets}
\end{figure}

\begin{table}
    \centering
    \begin{tabular}{l|c|r}
        \hline
        Doublet & $\lambda_1-\lambda_2$ [\AA] & $v$ [\kms] \\
        \hline
        C\,\textsc{iv} & $1548.20-1550.77$ & 497.2 \\
        Mg\,\textsc{ii} & $2796.35-2803.53$ & 768.7 \\
        Fe\,\textsc{ii} & $2374.46-2382.76$ & 1046.6 \\
        Fe\,\textsc{ii}' & $2586.65-2600.17$ & 1563.2 \\
        Al\,\textsc{iii} & $1854.72-1862.79$ & 1302.2 \\
        Si\,\textsc{iv} & $1393.76-1402.77$ & 1932.8 \\
        \hline
    \end{tabular}
    \caption{Doublets used in SB correlation function analysis.}
    \label{tab:doublets}
\end{table}

We found that the correlation function has a smooth component, which we model as follows:
\begin{equation}
    \xi_\mathrm{smooth} (v) = C + \frac{x_0}{1 + (v/v_0)^\gamma},
\end{equation}
where $C, x_0, v_0$ and $\gamma$ are fitting parameters. We then model the peaks as Gaussian functions:
\begin{equation}
    \xi_p(v) = A_p \mathrm{e}^{-(v-\mu_p)^2/2\sigma_p^2},
\end{equation}
where $A_p$ and $\sigma_p$ are free parameters, and the doublet separation is fixed to tabulated values: $\mu_p = c\ln (\lambda^p_2/\lambda^p_1)$. We fit for known metal doublets in Table~\ref{tab:doublets} by summing these functions: $\xi=\xi_\mathrm{smooth} + \sum_p \xi_p$. The residuals and their respective fits are in Fig.~\ref{fig:xi1d_sbboth_doublets}. This simple model accurately describes the important features in both correlation functions.

\begin{figure}
    \centering
    \includegraphics[width=\columnwidth]{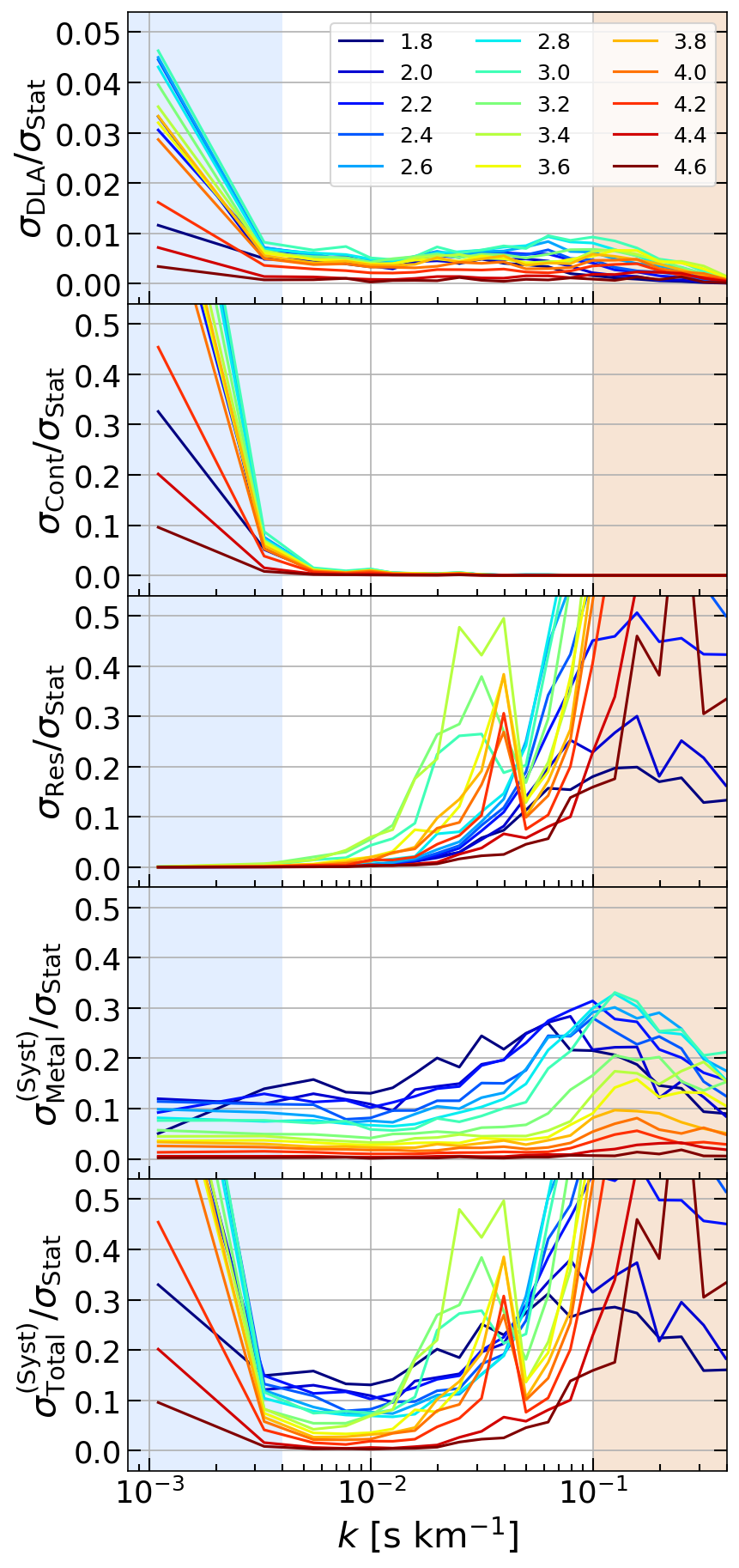}
    \caption{Systematic error estimates divided by the SB 1 subtracted bootstrap estimates. Continuum systematics affect the largest scales that are outside our conservative range. DLA systematics behave similarly, but they are an order of magnitude smaller. Metal systematics are relevant at low redshifts or high $k$.}
    \label{fig:syst_error_ratio}
\end{figure}

\subsection{Systematic error budget}
\label{subsec:systematics}
We identified four possible sources of systematic errors in our analysis, which are difficult to rectify by models. The continuum is fitted by hand and hard to reproduce. DLA removal is also partly manual in our measurement. For metals, no model at sufficient resolution exists.
Our systematic error budget is similar to \citet{chabanier_one-dimensional_2019} and based on the fiducial power parameters given in Section~\ref{sec:results}.
We provide each systematic error separately in our files, so that they can be scaled to other values readers see fit.
However, we recommend that these numbers are not modified for immediate cosmological fits. 
Our intention is to leave room for focused studies of each contaminant such as sub-DLAs, Lyman-limit systems (LLS), resolution modelling and metal line blending. 
These require dedicated studies; and the reader can adjust the error budget according to their findings.

\begin{itemize}
    \item \textbf{Resolution:} As we have discussed in Section~\ref{sec:data}, seeing conditions alter the resolution. 
    Due to this correction, we expect resolution accuracy to be smaller than normal.
    XQ-100 reports seeings with an average of 10\% precision. SQUAD however is much worse at 40\%.
    We then assign corresponding inaccuracies to $R$ values.
        \begin{equation}
            \sigma_{\mathrm{Res}} = r_m \times P_{\mathrm{fid}}(k) \times 2k^2\overline{R}^2,
        \end{equation}
    where $\overline{R}$ is the average resolution in a given redshift bin and $r_m$ is 0.1 for XQ-100 and 0.4 for KS.
    Note that even an undesirable 40\% change in resolution for SQUAD-like spectrum results in only 1--3\% change in \poned at $k=0.1$\skm. Our reported precision at this value is at least twice as large.
    Here, we assumed the resolving power is provided with perfect precision by the data sets, and ignored its contribution to the budget.

    \item \textbf{Metal:} There are two reasons to include a systematic error budget for metals: 1) Our statistical estimate of the metal power might be off. 2) Bootstrap error estimates might not reflect the truth. Assuming metal power is nearly constant with $k$ and $z$, we compared the fluctuations between redshifts and bootstrap error estimates. We found on average fluctuations between redshift bins are 6\% larger than the bootstrap error estimates. 
    Leaving room for the blending of metal lines we pick 10\% for our budget \citep{day_power_2019}.
    Using the fiducial power of the side band estimates, the systematic error is given by
        \begin{equation}
            \sigma_{\mathrm{Metal}}^\mathrm{(Syst)} = 0.10 \times P_{\mathrm{fid}}^{\mathrm{(SB1)}}(k).
        \end{equation}
        
    \item \textbf{Incomplete DLA removal:} Even though we used visual identification and catalogs, we leave room for missed DLAs or other high-density absorbers by assigning 1\% incompleteness to this possibility. We multiply the fiducial power estimate with redshift average of the absolute ratio of $P_{\mathrm{Kept}}/P_{\mathrm{Masked}}-1$ in Fig.~\ref{fig:dla_over_nodla} and DLA incompleteness ratio.
        \begin{equation}
            \sigma_{\mathrm{DLA}} = 0.01 \times P_{\mathrm{fid}}(k) \times \left\langle |P_{\mathrm{Kept}}/P_{\mathrm{Masked}}-1| \right\rangle_z 
        \end{equation}
    We note that the effects of sub-DLAs and LLS that are absent in catalogues can be larger than 1\% \citep{rogers_simulating_2018}.
    
    \item \textbf{Continuum:} We assign 10\% inefficiency to continuum marginalization and use the redshift average of the absolute ratio of $P_{\mathrm{No Mar}}/P_{\mathrm{Mar}}-1$ in Fig.~\ref{fig:cont_marg_ratio}. This error is also scaled with the fiducial power.
        \begin{equation}
            \sigma_{\mathrm{Cont}} = 0.10 \times P_{\mathrm{fid}}(k) \times \left\langle |P_{\mathrm{No Mar}}/P_{\mathrm{Mar}}-1| \right\rangle_z 
        \end{equation}
    Note that continuum errors themselves could be larger than 10\%. Our method however prevents these errors from contaminating \poned measurements by marginalizing out bulk of them,  namely a constant and a slope per chunk (not spectrum). 
    Here, our systematic budget really comes from the efficiency of this removal, which we assume to be 90\% effective.
    Furthermore, the modes that are most affected are not in our conservative range.
\end{itemize}

The results are summarized in Fig.~\ref{fig:syst_error_ratio}. 
DLA and continuum systematics affect the largest scales as expected, but DLA systematics are an order of magnitude smaller. 
Errors due to resolution inaccuracy becomes relevant near $k=0.1$\skm, but they do not overcome statistical errors.
Metal systematics are relevant at low redshifts.
In our conservative $0.004$\skm$<k<0.1$\skm and $z>1.8$  range, systematic errors are 19\% of the statistical errors on average.

\section{Discussion}
\label{sec:discussion}
\subsection{Statistical power of our results}
\label{subsec:warmdm}
We would like to compare the statistical power of our results to the existing measurements. In order to do so, we come up with a crude Fisher forecasting analysis, which replaces the actual \poned measurement with a fiducial power spectrum and only takes the covariance matrix into account. Our model is unquestionably primitive and ignores many complications including thermal state of the IGM. But it will serve as an adequate frame of reference.

We limit our simple forecast to a single parameter: the cut-off scale $k_\mathrm{cut}$. We assume a fiducial model as the "truth" for all data sets. Then, the error on $k_\mathrm{cut}$ is given by
\begin{equation}
    \frac{1}{\sigma^2_{\mathrm{cut}}} = \frac{\partial^2 \mathcal{L}}{\partial k_\mathrm{cut}^2} = \bm P_{,k_\mathrm{cut}}^T \mathbf{F} \bm P_{,k_\mathrm{cut}},
\end{equation}
where $\bm P_{,k_\mathrm{cut}}$ is analytically calculated and the Fisher matrix $\mathbf{F}$ is taken from actual measurements. We assume the following relative suppression transfer function for the 3D power spectrum.
\begin{equation}
    T^2(k) = [1+( k/k_\mathrm{cut})^{2}]^{-10}
\end{equation}
This expression comes from warm dark matter models as a fitting function \citep{bode_halo_2001, viel_constraining_2005}. We pick $k_\mathrm{cut}=100$\hmpc as our fiducial cut-off scale. We keep the modelling of Ly\,$\alpha$ forest $P_{\mathrm{3D}}$ out of scope of this estimation, but note that $P_{\mathrm{1D}}$ is highly sensitive to small scale physics including Jeans smoothing, reionization history \citep{hui_equation_1997, gnedin_probing_1998} and redshift space distortions. Instead, we use our fiducial fit in equation~(\ref{eq:pd13_fitting_fn}), and integrate by parts to find the suppression on $P_{\mathrm{1D}}$.
\begin{align}
    P_{\mathrm{1D}}(k; k_\mathrm{cut}) &= \int_k^{\infty} \frac{q\mathrm{d}q}{2\pi} P_{\mathrm{3D}}(q) T^2(k; k_\mathrm{cut})\\
    P_{\mathrm{3D}}(k) &= -\frac{2\pi}{k} \frac{\mathrm{d} P_{\mathrm{fit}}}{\mathrm{d}k}
\end{align}
Finally, we use Planck 2018 as our fiducial cosmology \citep{planck_collaboration_planck_2018-1}, and convert from distance units in Mpc\,$h^{-1}$ to velocity units in Ly\,$\alpha$ forest using $k_{\mathrm{Ly}\alpha} = k_{\mathrm{Mpc}} (1+z)/H(z)$, which roughly corresponds to a factor of 0.01.

% Note that a reliable error estimation has been a challenge for this small data sets, etc.

For \citet{irsic_lyman_2017}, we add systematic errors in quadrature to the diagonal of the covariance matrix and use all \poned points.
This yields $\sigma_{\mathrm{cut}}=32.9$\hmpc. 
% We note that \citet{irsic_new_2017} analysis incorporates some HIRES quasars in addition to XQ-100 only \citet{irsic_lyman_2017} results.
The same process yields $\sigma_{\mathrm{cut}}=20.1$\hmpc for \citet{chabanier_one-dimensional_2019}.
We use metal subtracted result from \citet{walther_new_2017} and limit to $k<0.1$\skm range. They do not provide a separate systematic error budget, so the covariance matrix is not modified. This yields $\sigma_{\mathrm{cut}}=19.0$\hmpc.
These numbers are also listed in Table~\ref{tab:sigma_mwdm}. 

For the statistical power of our measurement, we limit ourselves to a conservative $0.004$\skm$<k<0.1$\skm and $z>1.8$ range. 
This analysis with only statistical errors from bootstrap yields $\sigma_{\mathrm{cut}}=6.4$\hmpc for our results.
We then add our systematic error budget in quadrature to the diagonal of the covariance matrix, which increases the final error on $k_\mathrm{cut}$ to $\sigma_{\mathrm{cut}}=7.8$\hmpc.
Adding the largest scales $k<0.004$\skm does not improve this number as expected. 
However, $z=1.8$ bin brings this value down to 5.8\hmpc.
% However, extending high $k$ limit to 0.2\skm would bring this value down to 6.1\hmpc.
To summarize, we expect our results in conservative range to improve the cut-off scale sensitivity by a factor of 2.4.

Interestingly, including XQ-100 at the chunk level actually yields higher $\sigma_{\mathrm{cut}}=7.7$\hmpc with only statistical errors from bootstrap. This increased error from the bootstrap realizations reflects the resolution disparity between data sets.

\begin{table}
    \centering
    \begin{tabular}{l|r}
    \hline
         &  $\sigma_{\mathrm{cut}}$ [\hmpc]\\
         \hline
         This work (stat. only) & 6.4 \\
         This work (stat. + syst.) & 7.8 \\
         \citet{walther_new_2017} & 19.0\\
         \citet{chabanier_one-dimensional_2019} & 20.1 \\
         \citet{irsic_lyman_2017} & 32.9 \\
         \hline
    \end{tabular}
    \caption{Comparison of different data sets' statistical cut-off scale constraining power in terms of estimated errors, $\sigma_{\mathrm{cut}}$, for fiducial $k_\mathrm{cut}=100$\hmpc.  Without systematic error budget, our results can improve cut-off scale sensitivity by a factor of 3. Additional error budget decreases this factor to 2.4. }
    \label{tab:sigma_mwdm}
\end{table}

\subsection{Remaining challenges}
We assumed the noise is uncorrelated throughout, even though QMLE is capable of including pixel level correlations. 
Correlated noise has two effects. First, it changes the weighting in QMLE. This is not a big effect, since these weights roughly stay constant across $k$ bins. A second effect is the uncorrected spurious power. 
This is removed through the side band power subtraction, though not exactly because we combine data from different instruments. 
Furthermore, if the noise is correlated, it should be correlated at the pixel level, which corresponds to very small scales that are outside our conservative range. For example, $k=0.1$\skm corresponds to approximately 20 pixels for KODIAQ and SQUAD. 
Finally, our error estimates are from bootstraps, which will account for the correlated noise in the error bars. 
Therefore, we expect our \poned estimates not to be biased due to correlated noise.
However, the impact of correlated noise should be revisited as the precision of \poned increases in the future. 

One shortcoming of high-resolution quasar observations is that they are targeted for specific studies of e.g. strong absorbers or metallicity distribution. \citet{murphy_uves_2019} notes that some UVES quasars were specifically targeted due to known DLAs. The original goal of the KODIAQ survey was to study O\,\textsc{vi} absorption at $z>2.2$ \citep{omeara_second_2017}. Even though we mask or subtract contaminants, their cross correlations with the IGM remain. Therefore, unbiased high-resolution quasar observations are crucial to remedy the effect of this sampling bias.

The continuum errors remain a big challenge for \poned measurements at large scales in general, but especially for these types of quasar spectra. 
Our marginalization limits the shape of the errors to a constant and a slope, and is not fully descriptive given hand-fitted continua. 
One could imagine marginalizing out higher order terms and trying to find convergence. 
However, it is possible that this recipe is not convergent and extra degrees of freedom will eventually wipe out all information.
We defer a dedicated analysis to future work.
Another option would be finding and marginalizing out dominant terms in continuum by a PCA analysis.
Although such templates exist, they come with non-negligible errors with a range of 3--30\% \citep{suzuki_predicting_2005}.
This would further require an analytically well-defined, uniform continuum fitting procedure. 
In short, accurately describing the quasar continuum remains a much-needed but difficult task. 

A challenge for \poned analyses is estimating systematic error budget for DLAs and continuum errors. 
Our data is inhomogeneous, so it is additionally difficult to gauge. 
Our best understanding showed these errors contaminate low $k$ values.
We decided to remove these most contaminated bins from our measurement.
As precision increases with future surveys, rigorous description of DLA and continuum systematics (as well as their careful treatments) will be needed.
It is best if future surveys build these into their pipeline.

\section{Summary}
\label{sec:summary}
High-resolution, high-S/N quasar spectra can measure 1D Ly\,$\alpha$ forest power spectrum at smaller scales compared to large-scale structure surveys.  
At these scales, \poned is able to constrain the thermal state of the IGM and different dark matter models.

We applied the optimal quadratic estimator to the largest such data set by combining two publicly available data releases (KODIAQ + SQUAD) at the spectrum level, and performed a separate analysis for another publicly available data set XQ-100.
We then presented the weighted average of these two as our final \poned measurement, and found that it agreed with previous studies with reduced error bars.
We identified four systematic error sources for our analysis: incomplete DLA removal, inefficient continuum marginalization, resolution errors and metals.
As an advantage of the quadratic estimator, our method is not biased due to gaps and hence free from the respective systematic error.
These four systematics are scale and redshift dependent, but smaller than the statistical errors. 

Finally, to demonstrate the constraining power of this work, we performed a crude, single-parameter Fisher forecast analysis for the cut-off scale.
We estimate that our results are more sensitive to this cut-off scale by a factor of 2.4 than the previous measurements.

\section{Data Availability}
\label{sec:data_avail}
The data underlying this article are available in the article and in its online supplementary material.
Our quadratic estimator code is publicly available and can be downloaded from \url{https://bitbucket.org/naimgk/lyspeq}.
Our implementation of further reductions is a collection of personal scripts, and hence less documented. We also make these publicly available for reference. These can be found in \url{https://bitbucket.org/naimgk/qsotools}.

%%%%%%%%%%%%%%%%%%%%%%%%%%%%%%%%%%%%%%%%%%%%%%%%%%

\section*{Acknowledgements}
NGK and NP are supported by DOE, reference number DE-SC0017660. AFR acknowledges support by FSE funds trough the program Ramon y Cajal (RYC-2018-025210) of the Spanish Ministry of Science and Innovation.
VI is supported by the Kavli foundation.
PN acknowledges financial support from Centre National d'\'Etudes Spatiales (CNES).

Some of the data presented in this work were obtained from the Keck Observatory Database of Ionized Absorbers toward QSOs (KODIAQ), which was funded through NASA ADAP grants NNX10AE84G and NNX16AF52G along with NSF award number 1516777”.

This research is supported by the Director, Office of Science, Office of High Energy Physics of the U.S. Department of Energy under Contract No. 
DE-AC02-05CH11231, and by the National Energy Research Scientific Computing Center, a DOE Office of Science User Facility under the same 
contract; additional support for DESI is provided by the U.S. National Science Foundation, Division of Astronomical Sciences under Contract No. 
AST-0950945 to the NSF’s National Optical-Infrared Astronomy Research Laboratory; the Science and Technologies Facilities Council of the United Kingdom; the Gordon 
and Betty Moore Foundation; the Heising-Simons Foundation; the French Alternative Energies and Atomic Energy Commission (CEA); 
the National Council of Science and Technology of Mexico; the Ministry of Economy of Spain, and by the DESI 
Member Institutions.  The authors are honored to be permitted to conduct astronomical research on Iolkam Du’ag (Kitt Peak), a mountain with 
particular significance to the Tohono O’odham Nation.  



\bibliographystyle{mnras}
\bibliography{MyLibrary.bib} % if your bibtex file is called example.bib

\bsp	% typesetting comment
\label{lastpage}
\end{document}